\documentclass[preprints]{Definitions/mdpi} 

\renewcommand{\linenumbers}{}

\firstpage{1} 
\pubvolume{25}
\issuenum{9}
\articlenumber{1270}
\pubyear{2023}
\copyrightyear{2020}
\externaleditor{Academic Editor: Johan Anderson, \today}
\datereceived{1 August 2023} 
\dateaccepted{24 August 2023} 
\datepublished{29 August 2023} 
\hreflink{https://doi.org/10.3390/e25091270} 

\graphicspath{{./fig/}{./png/}}
\usepackage{bm}

\newcommand{\EQ}{\begin{equation}}
\newcommand{\EN}{\end{equation}}
\newcommand{\EQA}{\begin{eqnarray}}
\newcommand{\ENA}{\end{eqnarray}}

\newcommand{\EEq}[1]{Equation~(\ref{#1})}
\newcommand{\Eq}[1]{Equation~(\ref{#1})}
\newcommand{\Eqs}[2]{Equations~(\ref{#1}) and~(\ref{#2})}

\newcommand{\Sec}[1]{Section~\ref{#1}}

\newcommand{\Fig}[1]{Figure~\ref{#1}}

\newcommand{\Figp}[2]{Figure~\ref{#1}({#2})}
\newcommand{\Figsp}[3]{Figures~\ref{#1}({#2}) and ({#3})}

\newcommand{\Figs}[2]{Figures~\ref{#1} and \ref{#2}}

\newcommand{\Tab}[1]{Table~\ref{#1}}

\newcommand{\bra}[1]{\langle #1\rangle}
\newcommand{\bbra}[1]{\left\langle #1\right\rangle}

{}
{}

{}
{}

{}
{}
{}
{}
{}
{}
{}
{}
{}
{}
{}
{}
{}
{}

{}

{}

{}
{}
{}

\newcommand{\xxx}{\hat{\mbox{\boldmath $x$}} {}}

\newcommand{\xx}{\bm{x}}

\newcommand{\BB}{\bm{B}}

\newcommand{\EE}{\bm{E}}
\newcommand{\JJ}{\bm{J}}

\newcommand{\AAA}{\bm{A}}

\newcommand{\uu}{\bm{u}}

\newcommand{\nab}{{\bm{\nabla}}}

\newcommand{\SSSS}{\mbox{\boldmath ${\sf S}$} {}}

\newcommand{\AAAA}{\mbox{\boldmath ${\cal A}$} {}}

\newcommand{\DD}{{\rm D} {}}

\newcommand{\dd}{{\rm d} {}}

\newcommand{\const}{{\rm const}  {}}

\def\la{\mathrel{\mathchoice {\vcenter{\offinterlineskip\halign{\hfil
$\displaystyle##$\hfil\cr<\cr\sim\cr}}}
{\vcenter{\offinterlineskip\halign{\hfil$\textstyle##$\hfil\cr<\cr\sim\cr}}}
{\vcenter{\offinterlineskip\halign{\hfil$\scriptstyle##$\hfil\cr<\cr\sim\cr}}}
{\vcenter{\offinterlineskip\halign{\hfil$\scriptscriptstyle##$\hfil\cr<\cr\sim\cr}}}}}

\def\Pm{\mbox{\rm Pr}_{\rm M}}

\def\EEK{{\cal E}_{\rm K}}
\def\EEM{{\cal E}_{\rm M}}

\def\EEEl{{\cal E}_{\rm E}}

\def\cv{c_{\rm v}}

\def\cs{c_{\rm s}}

\def\vA{v_{\rm A}}

\def\vAz{v_{\rm A0}}

\def\epsK{\epsilon_{\rm K}}
\def\epsM{\epsilon_{\rm M}}

\def\Brms{B_{\rm rms}}
\def\Erms{E_{\rm rms}}

\def\half{\textstyle{\frac{1}{2}}}

\newcommand{\Gauss}{\,{\rm G}}

\Title{Electromagnetic Conversion into Kinetic and Thermal Energies}

\TitleCitation{Electromagnetic Conversion into Kinetic and Thermal Energies}

\Author{Axel Brandenburg $^{1,2,3,4,\dagger,\ddagger}$\orcidA{}, and Nousaba Nasrin Protiti$^{1,2}$\orcidB{}}

\AuthorNames{Axel Brandenburg and Nousaba Nasrin Protiti}

\AuthorCitation{Brandenburg, A.; Protiti, N.N.}

\address{%
$^{1}$ \quad Nordita, KTH Royal Institute of Technology and Stockholm University, Hannes Alfv\'ens v\"ag 12, 10691 Stockholm, Sweden; nopr1532@student.su.se \\
$^{2}$ \quad Oskar Klein Centre, Department of Astronomy, Stockholm University, AlbaNova, SE-10691 Stockholm, Sweden\\
$^{3}$ \quad School of Natural Sciences and Medicine, Ilia State University, 0194 Tbilisi, Georgia\\
$^{4}$ \quad McWilliams Center for Cosmology and Department of Physics, Carnegie Mellon University, Pittsburgh, Pennsylvania 15213, USA
}

\corres{Correspondence: brandenb@nordita.org}

\abstract{
Conversion of electromagnetic energy into magnetohydrodynamic energy
occurs when the electric conductivity changes from negligible to
finite values.
This process is relevant during the epoch of reheating in the early
Universe at the end of inflation and before the emergence of the
radiation-dominated era.
We find that the conversion into kinetic and thermal energies is primarily
the result of electric energy dissipation, while
magnetic energy only plays a secondary role in this process.
This means that, since electric energy dominates over magnetic energy
during inflation and reheating, significant amounts of electric energy
can be converted into magnetohydrodynamic energy when conductivity
emerges early enough, before the relevant length scales become stable.
}

\keyword{Electric energy; cosmological inflation; emergence of conductivity}

\begin{document}

\section{Introduction}

\begin{figure}[H]\begin{center}
\includegraphics[width=.70\textwidth]{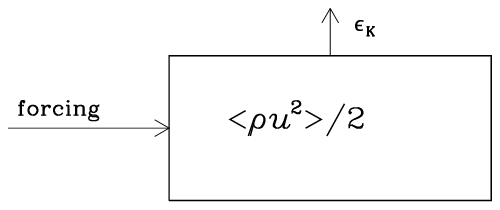}
\end{center}\caption[]{
Kinetic energy dissipation, $\epsK$, of forced turbulence with kinetic
energy density $\bra{\rho\uu^2}/2$, where $\rho$ is the density and $\uu$
is the velocity: in the steady state, everything that gets in does get out.
}\label{pboxes_hyd}\end{figure}

\begin{figure}[H]\begin{center}
\includegraphics[width=.70\textwidth]{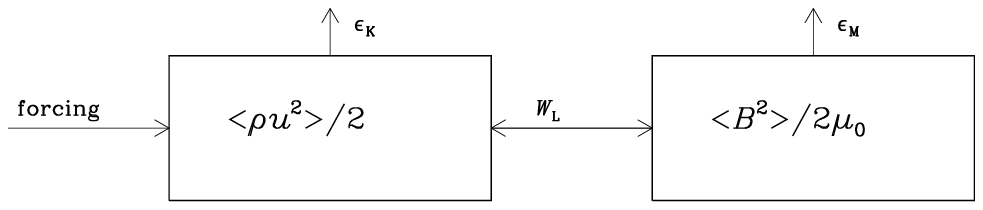}
\end{center}\caption[]{
Dissipation in dynamos: there are now two exit channels, $\epsK$ and
$\epsM$, and it is not clear who takes the lion's share.
Dynamo action corresponds to $W_{\rm L}<0$ (work done {\em against}
the Lorentz force, arrow to the right), although energy can also go
the other way around when an initial magnetic field decays.
}\label{pboxes_dyn}\end{figure}

In hydrodynamic turbulence, dissipation of energy is in principle
straightforward: it must be equal to the energy input accomplished
through forcing; see \Fig{pboxes_hyd}.
But when magnetic fields are involved, energy can be transferred
from kinetic energy to magnetic by performing work against the Lorentz
force, $W_{\rm L}$.
In that case, the situation is more complicated, because there are now
two exit channels, and it is {\em a priori} not clear, which of the two
takes the lion's share in specific situations; see \Fig{pboxes_dyn}.
A related problem may also occur when the electric energy reservoir is
involved, and especially when this energy reservoir is later absent due
to high conductivity.
Before getting to that, let us first recall the different situations
in hydrodynamic and hydromagnetic turbulence.

\Fig{pboxes_dyn} presumes that kinetic energy can be tapped by dynamo
action and converted into magnetic energy \citep{Mof78}.
This is a generic process that we now know works in virtually all types
of turbulent systems provided the electric conductivity is large enough
\citep{Tobias21}.
And here comes already the first problem.
Large conductivity means small magnetic diffusivity and therefore
also less dissipation \citep{Bra14}.
Looking at \Fig{pboxes_dyn} however, this seems puzzling:
In the steady state, the dynamo term $W_{\rm L}$ must be just as large
as the resistive term, $\epsM$.
Thus, if the dynamo is efficient, also the dissipation must be large,
which is not expected (and also not true).

The puzzle of efficient dynamo action, but inefficient dissipation
was solved by realizing that at large conductivity (and especially
large magnetic Prandtl number, which is the ratio $\Pm\equiv\nu/\eta$ of
kinematic viscosity $\nu$ to magnetic diffusivity $\eta$, is much larger
than unity), a second conversion occurs at smaller length scales where
magnetic energy can be converted back into kinetic energy.
This process was termed a reversed dynamo \citep{BR19}, and it happens
at small scales when $\Pm\gg1$.
The concept of a reversed dynamo was already introduced previously
in the context of large-scale flows in two-fluid systems driven by microscopic
fields and flows \cite{Mahajan+05}. However, in the context of Ref.~\citep{BR19},
the focus was on small-scale dynamos that drive small-scale flows by the
Lorentz force when $\Pm\gg1$.

While the conversion between magnetic and kinetic energies is
reasonably well understood, not much is known about the conversion from
electromagnetic energy, i.e., the sum of electric and magnetic energies,
into magnetic and hydrodynamic energies when the electric conductivity gradually increases.
Such a process is important at the end of cosmological inflation
\cite{FC23}.
A stochastic electromagnetic field may have been produced during
inflation and reheating \citep{Sharma+17}.
At the end of reheating, the electric conductivity of the Universe
increased.
As discussed in Ref.~\cite{BS21}, significant magnetic field losses can
occur if the increase in conductivity is slow, and especially when
the magnetic diffusivity is at an intermediate level for a long time.
In the two extreme cases of very large diffusivity (corresponding to a
vacuum with undamped electromagnetic waves), and very small diffusivity
(corresponding to nearly perfect conductivity), no significant losses
are expected.
It is only during the period when the magnetic diffusivity is at an
intermediate level that significant resistive losses can occur.


Once the conductivity has reached large values, i.e., when the magnetic
diffusivity is small, strong turbulent flows will be driven.
In that regime, the Faraday displacement current can be neglected and
the equations reduce to those of magnetohydrodynamics \cite{Alfven42}.
The resulting turbulent flows cause the magnetic field to undergo
turbulent decay with inverse cascading, as has been studied intensively
since the mid 1990s \citep{BEO96, CHB01, BJ04, Kahn+13, BKT15, BK17,
HS22}.
At some point around the time of recombination, the photon mean free path
becomes very large, and a process called Silk damping becomes important
\citep{Silk68}.
It results from the interactions between photons and the gas and damps out
all inhomogeneities in the photon--baryon plasma \citep{Kolb+Turner90}.
In Ref.~\cite{BEO97}, this was modeled as a strongly increased viscosity,
thereby making the magnetic Prandtl number even larger.
However, a more physical approach is to add a friction term of the form
$-\uu/\tau$ on the right hand side of the momentum equation \citep{BJ04}.
It is generally taken for granted that magnetic fields just survive
Silk damping without much additional loss, and that they are simply frozen
into the plasma.
However, the details of this process have not yet been modeled.
It is clear, however, that the assumption of a well-conducting Universe
is an excellent one, even after the epoch of recombination some 380,000
years after the Big Bang, when there were very few charged particles.
As we emphasize below, the electric conductivity was then still large
enough that the electric field was negligible---even in the voids
between galaxy clusters.
In cosmology, it is only near the end of inflation that the electric
field can play a significant role.
The electric energy density was then comparable to or in excess of the
magnetic energy density.

The goal here is to understand more quantitatively how much magnetic
energy survives during the conversions from electromagnetic fields to
magnetohydrodynamic fields as the conductivity increases.
We also consider, in more detail, the conversion from magnetic fields
to electric fields at the end of the cosmological reheating phase, which is when
both fields are still growing and not yet equal to each other---unlike
the situation when electromagnetic waves are already established and there is no longer any
growth.

\section{Energetics during the Emergence of Conductivity}

The evolution of the electric and magnetic fields, $\EE$ and $\BB$,
respectively, is given by the Maxwell equations, written here in SI units:
\begin{equation}
\frac{1}{c^2} \frac{\partial\EE}{\partial t}=\nab\times\BB-\mu_0\JJ,
\qquad\nab\cdot\EE=\rho_{\rm e}/\epsilon_0,
\label{dEdt}
\end{equation}
\begin{equation}
\frac{\partial\BB}{\partial t}=-\nab\times\EE,
\qquad\nab\cdot\BB=0,
\label{dBdt}
\end{equation}
where $c$ is the speed of light, $\mu_0$ is the vacuum permeability,
$\epsilon_0\equiv1/(\mu_0 c^2)$ is the vacuum permittivity,
and $\rho_{\rm e}$ is the charge density.
To close the equations, we use Ohm's law,
\begin{equation}
\JJ=\sigma\,(\EE+\uu\times\BB),
\label{Ohm0}
\end{equation}
where $\sigma$ is the electric conductivity and $\uu$ is the velocity.

In the very early Universe, inflation dilutes the plasma to the extent
that there are virtually no particles, and hence the electric conductivity
vanishes.
Eventually, a phase of reheating must have occurred.
One possibility is that the stretching associated with the cosmological
expansion leads to electromagnetic field amplification until the electric
field begins to exceed the critical field strength for the Schwinger
effect \cite{KA14} to lead to the production of charged particles,
and thereby to the emergence of electric conductivity.
This change in $\sigma$ implies the existence of a phase when $\sigma$
has an intermediate value for a certain duration.
This leads to a certain electromagnetic energy loss given by
$\JJ\cdot\EE$.
This is a well-known result in magnetohydrodynamics, where the
displacement current is ignored, so we have $\nab\times\BB=\mu_0\JJ$.
This is then used when deriving the magnetic energy equation by taking
the dot product of \Eq{dBdt} with $\BB$, so we have
\begin{equation}
\frac{\partial}{\partial t}\left(\BB^2/2\mu_0\right)
=-\BB\cdot\nab\times\EE/\mu_0
=\JJ\cdot\EE-\nab\cdot(\EE\times\BB/\mu_0),
\label{MagEnergy00}
\end{equation}
where we have introduced the Poynting vector $\EE\times\BB/\mu_0$.
In the following, we often adopt volume averaging, which we denote
by angle brackets.
They depend just on time $t$, but not on position $\xx$.
We also adopt periodic boundary conditions in all three directions,
so we call the domain triply periodic.
Since a divergence under triply-periodic volume averaging vanishes,
we just have
\begin{equation}
\frac{\dd}{\dd t}\bbra{\BB^2/2\mu_0}
=-\bbra{\JJ\cdot\EE}\quad\mbox{(ignoring the displacement current)}.
\label{MagEnergy0}
\end{equation}
The $\bra{\JJ\cdot\EE}$ term, in turn, has two contributions.
Using Ohm's law in the form
\begin{equation}
\EE=\JJ/\sigma-\uu\times\BB,
\label{Ohm3}
\end{equation}
we find $\bra{\JJ\cdot\EE}=\bra{\JJ^2}/\sigma-\bra{\JJ\cdot(\uu\times\BB)}$,
or, using $-\JJ\cdot(\uu\times\BB)=\uu\cdot(\JJ\times\BB)$, we have
\begin{equation}
\bra{\JJ\cdot\EE}=\bra{\JJ^2/\sigma}+\bra{\uu\cdot(\JJ\times\BB)},
\label{Ohm4}
\end{equation}
so part of the electromagnetic energy turns into Joule (or magnetic)
heating, $\epsM\equiv\bra{\JJ^2/\sigma}$, and another part is converted
into kinetic energy through work done by the Lorentz force,
$W_{\rm L}\equiv\bra{\uu\cdot(\JJ\times\BB)}$, which eventually also
becomes converted into heat through viscous (kinetic) heating, $\epsK$.
In the case of dynamo action discussed in the introduction, of course,
$W_{\rm L}$ is negative, so work is done {\em against} the Lorentz force.
This is why the direction of the arrow in \Fig{pboxes_dyn} is reversed.
Force-free magnetic fields have $W_{\rm L}=0$ and can therefore not be
sustained against dissipation, but they can be long lived if the current
density is small enough; see Ref.~\cite{Henriksen19} for examples.

\begin{figure}[H]\begin{center}
\includegraphics[width=\columnwidth]{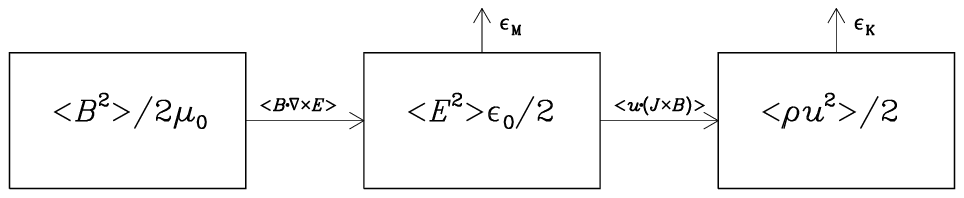}
\end{center}\caption[]{
Energy conversion from magnetic to kinetic energies via the
electric energy reservoir.
}\label{pboxes_BEU}\end{figure}

In the scenario where reheating is caused by the feedback from the
Schwinger effect, there would be thermal energy supply both from $\epsK$
and $\epsM$, leading therefore to a direct coupling between the resulting
heating and the emergence of $\sigma$.
The flows of energy between magnetic, electric, and kinetic energy
reservoirs is illustrated in \Fig{pboxes_BEU}.
We denote those by
\begin{equation}
\EEM\equiv\bra{\BB^2/2\mu_0},\quad
\EEEl\equiv\bra{\epsilon_0\EE^2/2},\quad\mbox{and}\quad
\EEK\equiv\bra{\rho\uu^2/2},
\label{EnergyDefs}
\end{equation}
respectively.
Their evolution equations can be obtained from \Eqs{dEdt}{dBdt},
along with the momentum and continuity equations,
\begin{equation}
\rho\frac{\DD\uu}{\DD t}=-\nab p+\JJ\times\BB+\nab\cdot(2\rho\nu\SSSS),
\label{Momentum}
\end{equation}
\begin{equation}
\frac{\DD\ln\rho}{\DD t}=-\nab\cdot\uu,
\label{Continuity}
\end{equation}
where $\DD/\DD t\equiv\partial/\partial t+\uu\cdot\nab$ is the advective
derivative, $p=\rho\cs^2$ is the pressure for an isothermal equation
of state with sound speed $\cs$, which is constant, $\nu$ is the viscosity, and
${\sf S}_{ij}=(\partial_i u_j+\partial_j u_i)/2-\delta_{ij}\nab\cdot\uu/3$
are the components of the rate-of-strain tensor $\SSSS$.

Note that, unlike the cases depicted in \Figs{pboxes_hyd}{pboxes_dyn},
there is no energy input in the system shown in \Fig{pboxes_BEU}.
This would change if we were to add forcing in the momentum equation in
\Eq{Momentum}.
We allude to this interesting possibility at the end of the conclusions
in \Sec{Conclusions}.
Another possibility that we do discuss in some detail is energy input
during the reheating phase at the end of inflation.
We turn to this aspect in \Sec{EnergyConversionsReheating}.

Taking the dot product of \Eq{Momentum} with $\uu$, using \Eq{Continuity},
integration by part, and the facts that $\partial_i u_j$ can be written
as the sum of a symmetric and an antisymmetric tensor, but that the
multiplication with $\SSSS$ (a symmetric and trace-free tensor) gives
no contribution when $\delta_{ij}\nab\cdot\uu/3$ is added, we find
that ${\sf S}_{ij}\partial_i u_j=\SSSS^2$, and thus obtain the evolution
equation for the kinetic energy in the form
\begin{equation}
\frac{\dd}{\dd t} \bbra{\rho\uu^2/2}=-\bbra{\uu\cdot\nab p}
+\bbra{\uu\cdot(\JJ\times\BB)}-\bbra{2\rho\nu\SSSS^2},
\label{EEKeqn1}
\end{equation}
which we can also write more compactly as
$\dot{\cal E}_{\rm K}=W_{\rm P}+W_{\rm L}-\epsK$,
where $W_{\rm P}=-\bra{\uu\cdot\nab p}$
has been defined as the work done by the pressure force,
$\epsK=\bra{2\rho\nu\SSSS^2}$ is the viscous heating, 
and the dot on the kinetic energy ${\cal E}_{\rm K}$
(not to be confused with $\epsK$) denotes a time derivative.
Here we have made use of the fact that the divergence
$\nab\cdot(p\uu)=\uu\cdot\nab p+p\nab\cdot\uu$ has a vanishing
volume average for a triply-periodic domain, and therefore
$-\bra{\uu\cdot\nab p}=\bra{p\nab\cdot\uu}$, making it clear that this
term leads to compressional heating and was found to be important in
gravitational collapse simulations \cite{BN22}.
We will see later that, when energy is supplied through $W_{\rm L}$,
the energy is used to let the kinetic energy grow ($\dot{\cal E}_{\rm K}>0$)
and to drive viscous heating, i.e., we have
\begin{equation}
W_{\rm L}=\dot{\cal E}_{\rm K}+\epsK-W_{\rm P}.
\label{EEKeqn3}
\end{equation}
We recall that the dot on $\EEK$ denotes a time derivative.
The term $W_{\rm P}$ is usually small and negative and thus also
contributes (but only little) to increasing thermal energy.
In the present simulations, we used an isothermal equation of state and
thus ignored the evolution of thermal energy, ${\cal E}_{\rm T}=\bra{\rho e}$,
where $e=\cv T$ is the internal energy, $\cv$ is the specific heat at
constant volume, and $T$ is the temperature.
If we had included it, we would have had
\begin{equation}
\dot{\cal E}_{\rm T}=\epsM+\epsK-W_{\rm P}.
\label{EETeqn}
\end{equation}
This thermal evolution is important in simulations of thermal
magneto-convection \citep{Bra+96}, where it facilitates buoyancy
variations, or in simulations of the magneto-rotational instability, where
potential energy is converted into kinetic and magnetic energies
that then dissipate as heat and radiation \cite{BNST95}.
For our purposes, however, it suffices to integrate instead
the kinetic and magnetic contributions in time, i.e., to compute
$\int\epsK\,\dd t$ and $\int\epsM\,\dd t$, respectively.

Let us now discuss the interplay between electric and magnetic energies.
This interplay is usually ignored in magnetohydrodynamics, where the
evolution of the electric field, i.e., the Faraday displacement current,
is ignored \cite{Alfven42}.
Taking the dot product of \Eq{dEdt} with $\EE/\mu_0$ and using
$1/(\mu_0 c^2)=\epsilon_0$, we obtain
\begin{equation}
\frac{\partial}{\partial t}\left(\epsilon_0\EE^2/2\right)=
\frac{\EE}{\mu_0 c^2}\cdot\frac{\partial\EE}{\partial t}=
\EE\cdot\nab\times\BB/\mu_0-\JJ\cdot\EE,
\label{dE2dt}
\end{equation}
so, after averaging, we have
\begin{equation}
\frac{\dd}{\dd t}\bbra{\epsilon_0\EE^2/2}=
\bbra{\EE\cdot\nab\times\BB/\mu_0}-\bbra{\JJ\cdot\EE}.
\label{dE2dt2}
\end{equation}
Next, taking the dot product of \Eq{dBdt} with $\BB/\mu_0$, we obtain
\begin{equation}
\frac{\partial}{\partial t}\left(\BB^2/2\mu_0\right)=-\BB\cdot\nab\times\EE/\mu_0.
\label{MagEnergy1}
\end{equation}
In view of the $\bra{\EE\cdot\nab\times\BB/\mu_0}$ term in \Eq{dE2dt2},
it is convenient to rewrite \Eq{MagEnergy1} in the form
\begin{equation}
\frac{\partial}{\partial t}\left(\BB^2/2\mu_0\right)=-\EE\cdot\nab\times\BB/\mu_0
-\nab\cdot(\EE\times\BB/\mu_0).
\label{MagEnergy2}
\end{equation}
Again, given that the Poynting flux divergence vanishes under a
triply-periodic volume averaging, we have
\begin{equation}
\frac{\dd}{\dd t}\bbra{\BB^2/2\mu_0}=-\bbra{\EE\cdot\nab\times\BB}/\mu_0.
\label{MagEnergy3}
\end{equation}
Note here the difference to \Eq{MagEnergy0}, which ignores the
displacement current.
An equation similar to \Eq{MagEnergy0} can only be recovered for the
sum of electric and magnetic energies, which yields
\begin{equation}
\frac{\dd}{\dd t}
\left(\bbra{\BB^2/2\mu_0}+\bbra{\epsilon_0\EE^2/2}\right)
=-\bbra{\JJ\cdot\EE}.
\label{MagEnergy4}
\end{equation}
An important property of well-conducting media that are considered in
magnetohydrodynamics is that the electric energy is negligible compared
to the magnetic energy.
In that limit, \Eqs{MagEnergy0}{MagEnergy4} do indeed become equivalent.

More compactly, we can then write \Eq{MagEnergy3} in the form
$\dot{\cal E}_{\rm M}=-Q_{\rm E}$, where
$Q_{\rm E}=\bra{\EE\cdot\nab\times\BB}$ acts as a source in
$\dot{\cal E}_{\rm E}=Q_{\rm E}-\epsM-W_{\rm L}$.
Thus, we clearly see that the electric energy reservoir is not a secondary
one whose energy content is small because of inefficient coupling, but
it is an unavoidable intermediate one through which magnetic energy is
channeled efficiently further to kinetic and thermal energies.
This raises the question how safe is the neglect of the displacement
current when prior to the emergence of conductivity the electric energy
dominates over magnetic.
This is a typical situation in inflationary magnetohydrodynamics scenarios
that we consider later in this paper.
Before that, we first discuss the nonconducting case where electric and
magnetic energy densities are equally large.

\section{Numerical Experiments with Different Temporal Conductivity Variations}

To illustrate the conversion from electromagnetic energy to
magnetohydrodynamic and thermal energies during the emergence of electric
conductivity, let us consider here a simple one-dimensional experiment.

\subsection{Electromagnetic Waves and Their Suppression by Conductivity}

In one dimension with $\partial/\partial x\neq0$ and $\sigma=0$,
we can have electromagnetic waves, for example
$B_y^\pm(x,t)=B_0 \sin k(c\mp ct)$ and
$E_z^\pm(x,t)=\mp k B_0\sin k(c\mp ct)$, traveling in the
positive (negative) $x$ direction.
Note that the electric and magnetic energies are here equal to each
other.
However, when $\sigma$ becomes large, $|\EE|$ becomes suppressed.
To understand this suppression, let us look at \Eq{dEdt}.
When $\sigma$ becomes large, the $\nab\times\BB$ term no longer needs
to be balanced by the displacement current, but by the actual current.
Inserting $\JJ=\sigma\EE$ (for the comoving current density), we find
$\EE=\eta\nab\times\BB$, so we expect $|\EE|/|c\BB|=O(\eta k/c)$.
Thus, once $\sigma$ becomes large, $|\EE|/c$ becomes suppressed relative
to $|\BB|$ by a factor, $\eta k/c=k/(c\mu_0\sigma)$.
Additionally, as mentioned above, both $|\EE|$ and $|\BB|$ become
suppressed due to the intermediate phase when $\sigma$ is neither small
nor large yet.
This was discussed in the appendix of Ref.~\cite{BS21},
who found that for a linearly increasing conductivity profile
$\sigma(t)=\sigma_{\max}\,t/t_{\rm trans}$ during a certain time interval
$t_0\leq t\leq t_0+t_{\rm trans}$ of duration $t_{\rm trans}$ and starting
at $t=t_0$, there was an amplitude drop, whose
value increases approximately inversely proportional to
$\eta_{\min} k^2t_{\rm trans}$, where $\eta_{\min}=1/\mu_0\sigma_{\max}$.

To specify the temporal variation of the conductivity profile, we define
a piecewise linear function that goes from 0 (for $t\leq t_0$) to 1
(for $t\geq t_0+\tau_0$) through
\begin{equation}
\Theta=\max\left[\min\left(\frac{t-t_0}{\tau_0},\,1\right),\,0\right].
\end{equation}
The linear $\sigma$ profile used in Ref.~\cite{BS21} is given by
\begin{equation}
\sigma(t)=\sigma_{\min}+(\sigma_0-\sigma_{\min})\,\Theta(t),
\end{equation}
where $\sigma_0=1/\mu_0\eta_0$ and $\sigma_{\min}=1/\mu_0\eta_{\max}$.
Here, we also study a profile whose logarithm is linearly varying.
We therefore refer to it as a logarithmic profile, which is of the form
\begin{equation}
\sigma(t)=\sigma_0\exp\left\{\ln(\sigma_{\max}/\sigma_0)\,
\left[1-\Theta(t)\right]\right\},
\end{equation}
which allows us to specify the duration, during which $\sigma$ transits
by an order of magnitude for any value of $\sigma$.
For the linear $\sigma$ profile, by contrast, the duration would be
different for different $\sigma$ ranges, and it would be very short for
large values of $\sigma$.

For the simulations, we use the {\sc Pencil Code} \cite{JOSS}, which
employs the magnetic vector potential $\AAA$, so that $\BB=\nab\times\AAA$
is always divergence-free.
The evolution equation for $\AAA$ can then be written as
\begin{equation}
\frac{1}{c^2} \frac{\partial^2\AAA}{\partial t^2}-\nabla^2\AAA+\nab\nab\cdot\AAA
+\frac{1}{\eta(t)} \left(\frac{\partial\AAA}{\partial t}+\uu\times\BB\right)=0,
\label{FullAeqn}
\end{equation}
where $\nab\cdot\AAA=0$ if the Coulomb gauge is used.
In the {\sc Pencil Code}, different gauges are possible, but the Weyl
gauge with $\partial\AAA/\partial t=-\EE$ is the one used in that code.
In that case, the $\nab\nab\cdot\AAA$ term must be retained.
\EEq{FullAeqn} shows that in a vacuum, where $1/\eta\to0$, one recovers
a standard wave equation for waves with propagation speed $c$.
In the opposite limit, where $\eta\to0$, one can neglect the
$(\eta/c^2)\,\partial^2\AAA/\partial t^2$ term and one recovers the usual
induction equation, where $\eta\nabla^2\AAA$ acts as a diffusion term.

\subsection{Transition to the High-Conductivity Regime for Different Parameters}
\label{Transition}

The transition to the high-conductivity regime involves the conversion of
electromagnetic waves to magnetohydrodynamic waves \cite{Alfven42}.
One can imagine that this process is more efficient when the frequencies
of both waves are equal.
In the high conductivity regime, the frequency of magnetohydrodynamic
waves depends on the strength of the imposed magnetic field, $B_0$,
which determines the nominal Alfv\'en speed, $\vAz=B_0/\sqrt{\rho\mu_0}$.
Since Alfv\'en waves propagate along the magnetic field, and since
$\partial/\partial x\neq0$, we impose the magnetic field also in the
$x$ direction, i.e., we write $\BB=\xxx B_0+\nab\times\AAA$, where
$\nab\times\AAA$ is the departure from the imposed magnetic field.
In the following numerical experiments, we choose $t_0=0$.

\begin{figure}[t!]\begin{center}
\includegraphics[width=.7\textwidth]{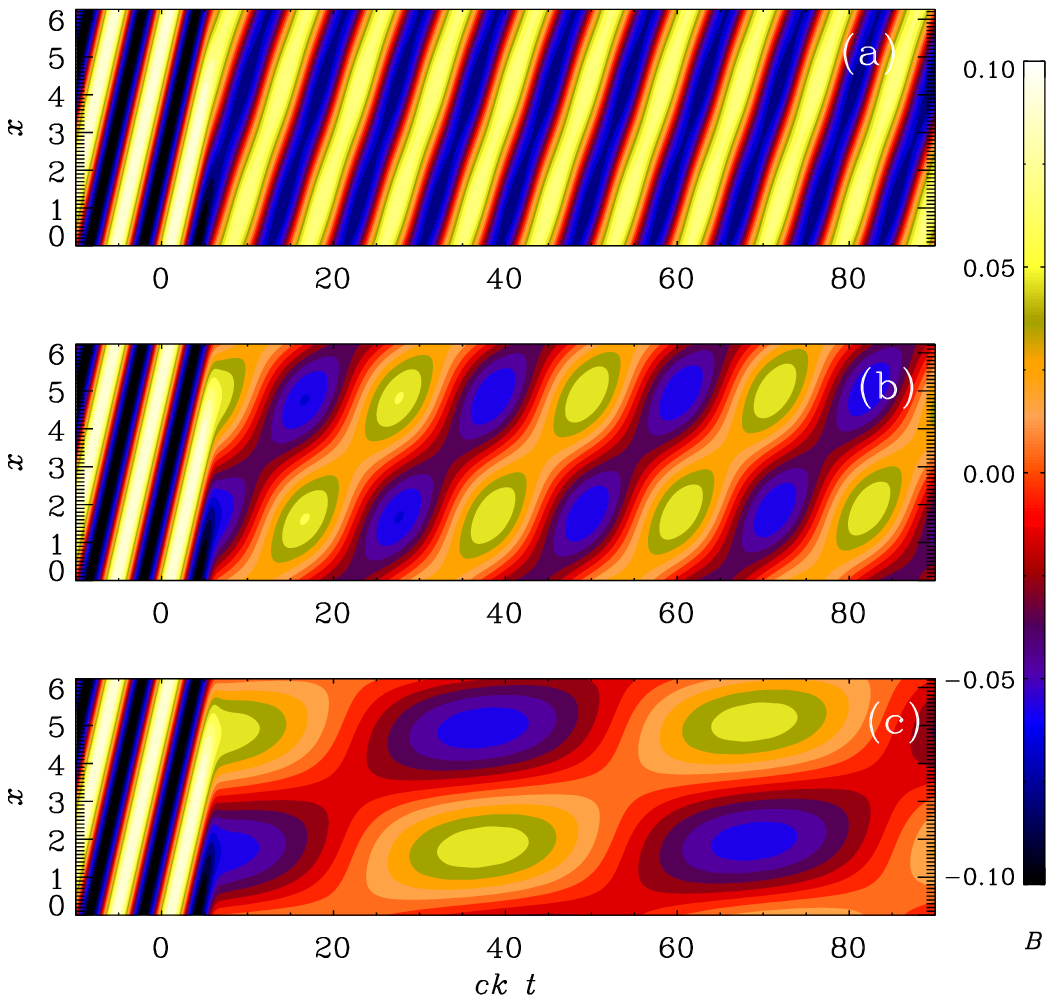}
\end{center}\caption[]{
Evolution of $B_y(x,t)$ for the logarithmic $\sigma$ profile
with (a) $\vAz=1$, (b) $\vAz=0.3$, and (c) $\vAz=0.1$, and $t_{\rm trans}=10$
in all cases.
}\label{pmaps}\end{figure}

The result is shown in \Fig{pmaps}, where we compare $B_y(x,t)$ as a
colored contour plot in the $xt$ plane.
We compute the solution in a domain of size $L$, so the lowest wave
number is $k_1=2\pi/L$.
The density is initially uniform and equal to $\rho_0$.
In the following, we use units where $c=k_1=\rho_0=1$.
We see that for $\vAz=1$, the wave propagates almost unaffectedly by
the switch to high conductivity.
Here, the frequency of the electromagnetic wave is $ck_1=1$ and the
nominal frequency of the Alfv\'en wave, $\vAz k$, is also unity but the
actual frequency is slightly less than that.
This is because of special relativity effects forcing the wave speed to
be always less than $c$.
In fact, the actual wave speed is $\vA=\vAz/(1+\vAz^2/c^2)^{1/2}$
\cite{Gedalin93}.


\begin{figure}[t!]\begin{center}
\includegraphics[width=.7\textwidth]{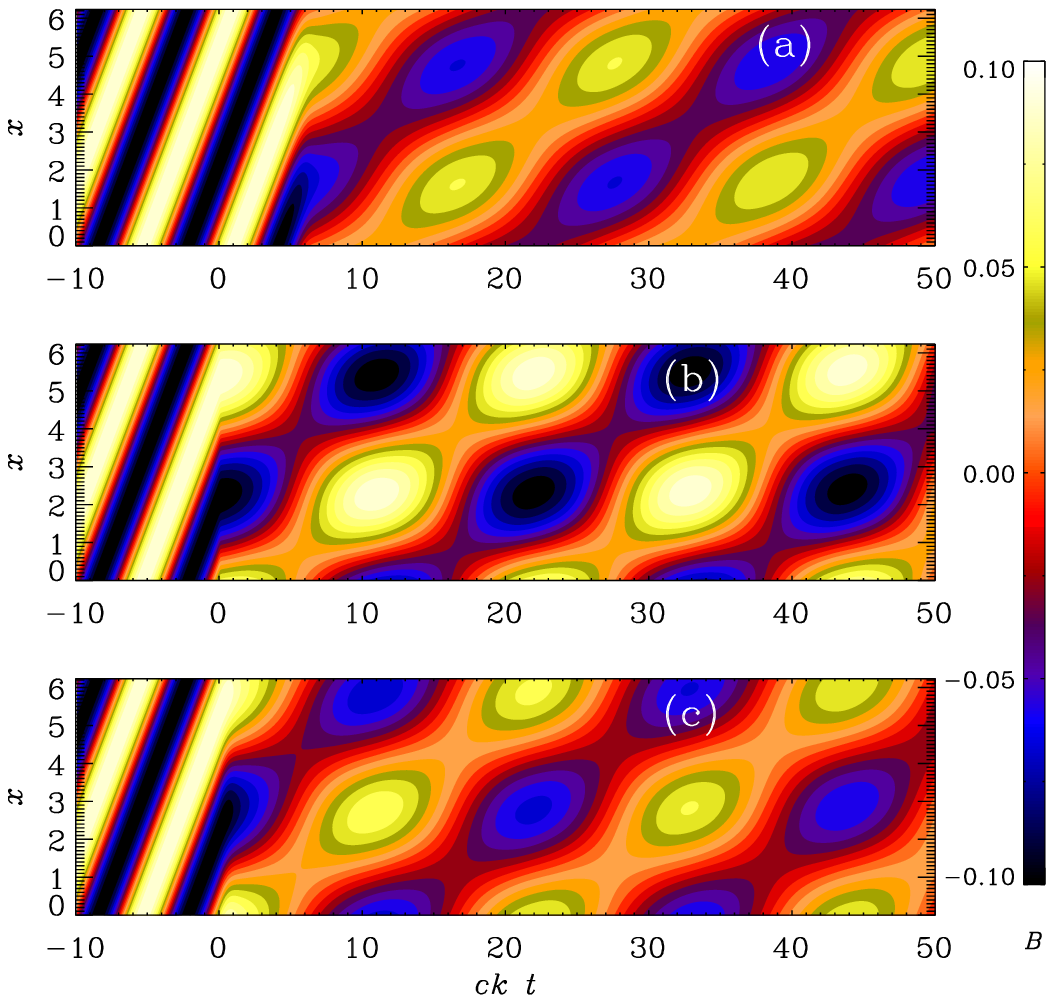}
\end{center}\caption[]{
Evolution of $B_y(x,t)$ for the logarithmic $\sigma$ profile
with $\vA=0.3$ and (a) the logarithmic $\sigma$ profile with $t_{\rm trans}=10$,
(b) the linear $\sigma$ profile with $t_{\rm trans}=10$, and
(c) the linear $\sigma$ profile with $t_{\rm trans}=500$.
}\label{pmaps03}\end{figure}

For smaller values of $\vAz$, we see that not only the wave speed
is less, as seen from the shallower inclination of the pattern in
\Figsp{pmaps}{b}{c}, but there is also a certain drop of the wave
amplitude, and there is also an additional modulation resulting from an
effective initial condition at $t=0$ that does not match the eigenfunction
for an Alfv\'en wave.

In \Fig{pmaps03}, we compare the logarithmic $\sigma$ profile with the
linear one using $\vAz=0.3$.
For the logarithmic profile, the drop in amplitude is clearly larger
than that for the linear $\sigma$ profile.
To obtain a similar drop with the linear $\sigma$ profile, one would
need to increase $t_{\rm trans}$ to about 500; see \Figp{pmaps03}{c}.

\begin{figure}[t!]\begin{center}
\includegraphics[width=.7\textwidth]{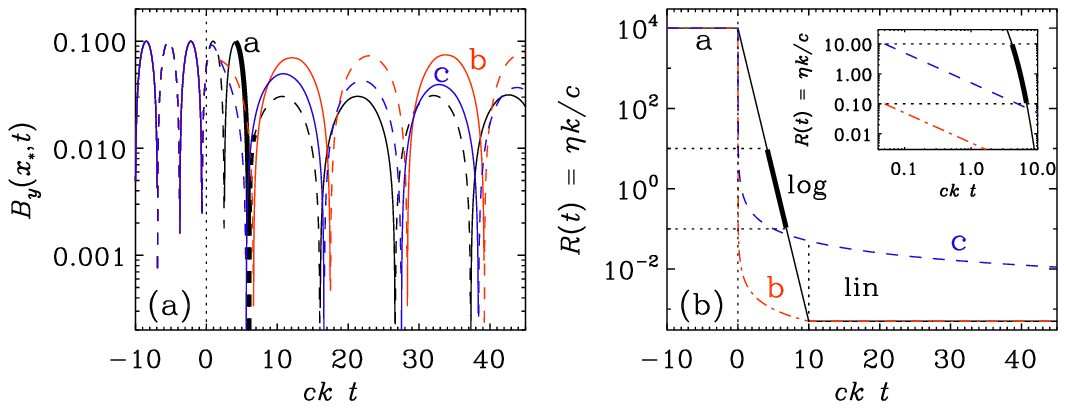}
\end{center}\caption[]{
(a) Evolution of $B_y$ at one specific point $x=x_*$ in the three
runs of \Figsp{pmaps03}{a}{c}.
Note that the drop of the wave amplitude after $t=0$, and
specifically at $t=50$, is similar for runs \textbf{a} and \textbf{c},
but much less for \textbf{b}.
(b) Dependence of the nondimensional resistivity $R(t)=\eta(t)
k/c$ for the logarithmic profile with $t_{\rm trans}=10$ in run \textbf{a}
(black), and the linear profile with $t_{\rm trans}=10$ in run \textbf{b}
(red) and $t_{\rm trans}=500$ in run \textbf{c} (blue).
The inset shows a blow-up of a narrow strip around $R=1$ using
a logarithmic time axis.
We see from the inset that the time spent in $R(t)$ traversing unity
by a margin of one order of magnitude (marked by the thick part of the
black line) is similar for \textbf{a} and \textbf{c}, but virtually
non-existing for \textbf{b}.
}\label{pvid_last_D128e}\end{figure}

To be more quantitative, we compare in \Figp{pvid_last_D128e}{a}
the evolution of $B_y$ at one specific point $x=x_*$ for the three
runs of \Figsp{pmaps03}{a}{c}.
Note that the drop of the wave amplitude after $t=0$ is similar
for runs \textbf{a} and \textbf{c}, but much less for run~\textbf{b}.


In \Figp{pvid_last_D128e}{b}, we also show how $\sigma$ varies.
We do this by plotting the nondimensional resistivity
\begin{equation}
R(t)\equiv\eta(t) k/c,
\end{equation}
which decreases from $10^4$ to $5\times10^{-4}$.
We recall that it is also this ratio that we identified in the beginning
of this section as the one that characterizes the value of $|E|/|cB|$.
We see that most of the decay happens when it transits through unity.
Owing to the logarithmic nature of the profile, the quantity $R(t)$
spends a time interval of about $ck t_{\rm trans}=5$ while $R(t)$ changes from
10 to 0.1.
By contrast, for the linear profile, the time interval is virtually
non-existing.
For $t_{\rm trans}=500$, on the other hand, $ck t_{\rm trans}$ is similar to
what led to the to a similar decay for the logarithmic profile.
This is also confirmed by the inset of panel \textbf{b}, which shows
that $R(t)$ traverses unity by a margin of one order of magnitude
for \textbf{a} and \textbf{c}, but not for \textbf{b}.
The results discussed above confirm that the relevant time interval is
indeed that where $R(t)$ is within an order of magnitude around unity.

\begin{figure}[H]\begin{center}
\includegraphics[width=.7\textwidth]{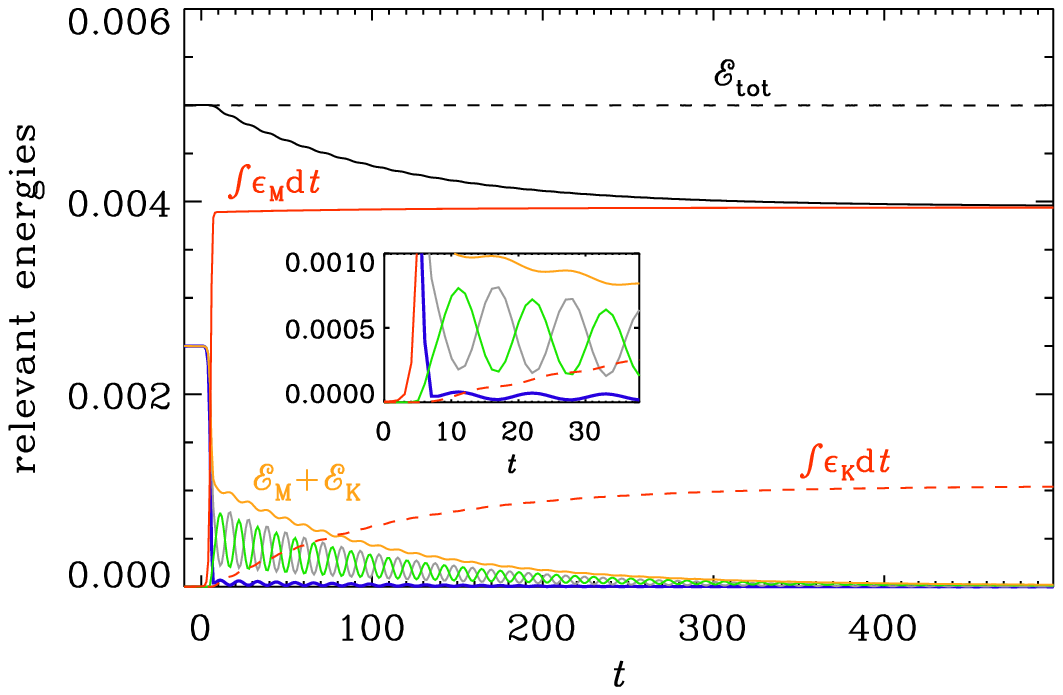}
\end{center}\caption[]{
Initially, all the energy is in electromagnetic energy,
$\EEEl+\EEM$ for $\nu=0.01$ and $\eta_{\rm fin}=5\times10^{-4}$.
In the end, all the energy is converted into heat.
The red lines give the integrated Ohmic and viscous energy gains,
$\int\epsM\,\dd t$ and $\int\epsK\,\dd t$, respectively.
At intermediate times, this energy is distributed to equal amounts
among kinetic energy $\EEK$ (green lines) and magnetic energy $\EEM$
(gray lines).
The orange lines shows their sum, $\EEK+\EEM$.
The blue lines represent $\EEEl$.
The inset shows a blow-up of the same graph around the origin.
We see that $\EEEl$ varies in phase with $\EEK$, but an anti-phase both
with $\EEM$ and the residual $\EEM+\EEK$.
}\label{pdec_D128j_T10a_etaE0_nu1em2}\end{figure}

Looking at \Fig{pdec_D128j_T10a_etaE0_nu1em2}, we see that the electric
energy was initially equal to the magnetic one, but as the conductivity
increases, there is a rapid decline of electric energy ($\dot{\EEEl}<0$),
and most of it dissipates thermally, while only a small fraction
($<10\%$ for $t_{\rm trans}=10$) is transferred to kinetic energy.
It turns out that the mean magnetic and electric energy densities decay
like $\exp(-\nu k^2 t)$, i.e., without a factor of $2$ in the exponent,
reflecting therefore not a change in the kinetic energy, but rather in
the velocity, which enters through the work term $W_{\rm L}$.).
Furthermore, the ratio is here $\EEEl/\EEM\approx10$ for $\Pm=20$.
Note that the oscillations in $\EEM+\EEK$ (orange lines) are
compensated mostly entirely by those in $\EEEl$ (blue lines).

\begin{figure}[t!]\begin{center}
\includegraphics[width=.7\textwidth]{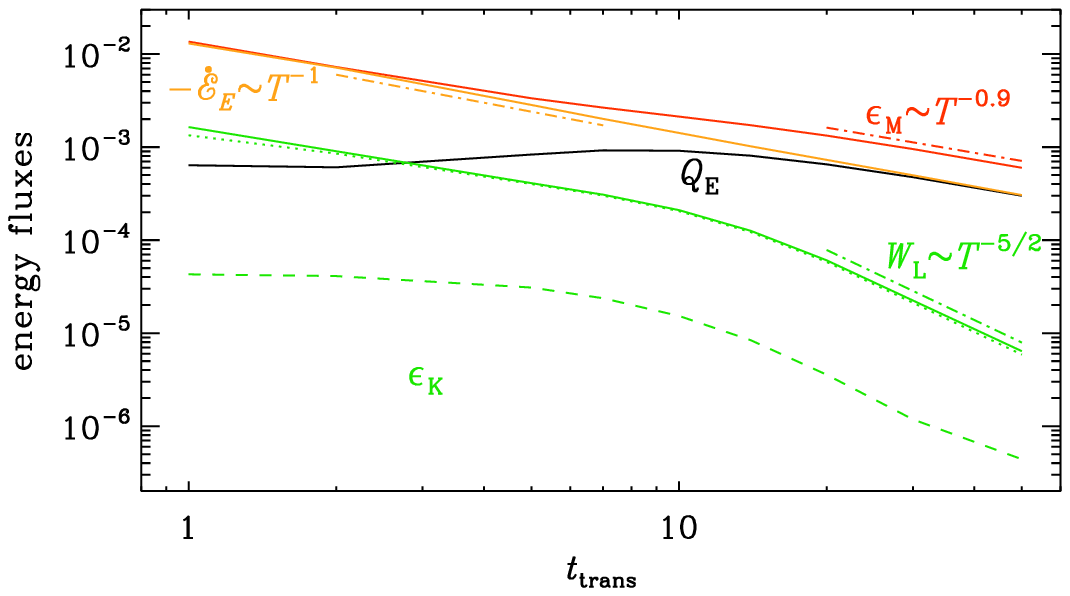}
\end{center}\caption[]{
Evolution of energy fluxes for the model with
the logarithmic conductivity profile with
$\eta=5\times10^{-4}=\nu$ at late times.
In all cases, the initial diffusivity is $\eta_{\rm ini}=10^4$.
The main difference to the run with a larger viscosity
is that $\epsK$ is larger.
}\label{presults_Tdep}\end{figure}

In \Fig{presults_Tdep}, we show the evolution of various
energy fluxes.
We see that the magnetic energy decays and gives off energy
to the electric energy reservoir through the term
$Q_{\rm E}=\bra{\EE\cdot\nab\times\BB}>0$.
The magnetic heating is thus composed of the following terms:
\begin{equation}
\epsM=-\dot{\cal E}_{\rm E}+Q_{\rm E}-W_{\rm L}.
\label{epsM_eqn}
\end{equation}
For rapid transits, $t_{\rm trans}\la5$, $Q_{\rm E}$ is small compared
with $-\dot{\cal E}_{\rm E}$, so $\epsM$ is mostly entirely the result
of exhausting electric energy, i.e., $\dot{\cal E}_{\rm E}<0$.
For longer transits, $t_{\rm trans}>10$,
$Q_{\rm E}\approx-\dot{\cal E}_{\rm E}$, so $\epsM$ is supplied
to about 50\% through $Q_{\rm E}$ and to another 50\% through
$-\dot{\cal E}_{\rm E}$.
These differences are summarized in \Tab{Tregimes}.

\begin{specialtable}[t!]\caption{
The two regimes of energy transfer for short and long transits.
\label{Tregimes}}
\begin{tabular}{lll}
\toprule
& rapid transits & long transits \\
\midrule
criterion & $t_{\rm trans}<10$ & $t_{\rm trans}>10$ \\
Lorentz work & $W_{\rm L}/\epsM\approx0.1$ & $W_{\rm L}/\epsM\approx5\,t_{\rm trans}^{-1.6}$ \\
heating      & $\epsM\approx-\dot{\cal E}_{\rm E}$ and $Q_{\rm E}\ll\epsM$
             & $\epsM\approx-0.5\,\dot{\cal E}_{\rm E}\approx0.5\,Q_{\rm E}$ \\
\bottomrule
\end{tabular}
\end{specialtable}

\begin{figure}[t!]\begin{center}
\includegraphics[width=.7\textwidth]{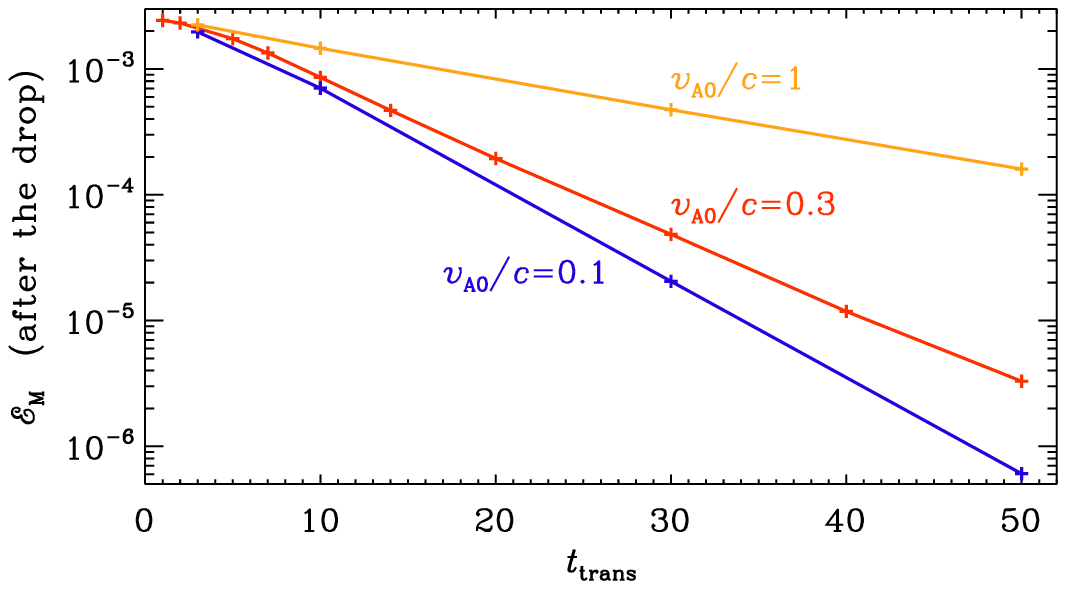}
\end{center}\caption[]{
$\EEM$ at $t=100$, i.e., after the conductivity
has increased to large value, vs $t_{\rm trans}$
for $\vAz/c=1$ (orange), $\vAz/c=0.3$ (red), and
$\vAz/c=0.1$ (blue).
}\label{presults_Tdep_energies}\end{figure}

In connection with \Fig{pmaps}, we noted that there is a certain drop
of the wave amplitude after the transit to large conductivity.
This drop was larger for a larger ratio of the electromagnetic to Alfv\'en
wave speeds.
When the nominal Alfv\'en speed was equal to the speed of light, the
drop was small.
In \Fig{presults_Tdep_energies} we quantify this by plotting $\EEM$
at $t=100$, i.e., after the conductivity has increased to a large value,
vs $t_{\rm trans}$ for different values of $\vAz/c$.
We confirm the results of Ref.~\cite{BS21} where the logarithmic
drop was found to depend linearly on the value of $t_{\rm trans}$.
However, we now also see that the slope of this curve decreases with
increasing Alfv\'en wave speed.

\section{Cosmological Application Prior to Radiation Domination}

As alluded to above, the end of inflation might provide an opportunity
to illustrate electromagnetic energy conversion, because in that case,
the electric energy can greatly exceed the magnetic one.

\subsection{Magnetic Fields in Cosmology}

In the present Universe, magnetic fields are constantly being regenerated
by dynamo action on all scales up to those of galaxy clusters.
The energy source is here gravitational, which is released through
accretion or direct collapse.
Magnetic fields may also be present on even larger scales.
However, in the locations between galaxy clusters, i.e., in what is
often referred to as voids, it is generally thought impossible to produce
magnetic fields through contemporary dynamo action; see
Refs.~\cite{DN13, Sub16} for reviews on the subject.
Nevertheless, indirect evidence for the existence of magnetic fields
in voids, and more specifically for lower limits of the magnetic field
strength, comes from the non-observation of secondary photons in the
halos to blazars, which are active galactic nuclei producing TeV photons.
These photons interact with those of the extragalactic background light
through inverse Compton scattering to produce GeV photons.
Those secondary GeV photons are not observed.
Their non-observation could be explained by an intervening magnetic field
of about $10^{-16}\Gauss$ on a megaparsec scale \cite{NV10, Taylor+11}.
This field would deflect electrons and positrons in opposite directions,
preventing them from recombining and thereby disrupting the energy
cascade toward the lower GeV photons.

The non-observation of GeV photons might have other reasons, for
example plasma instabilities that disrupt the electron--positron beam
\cite{Broderick+12, Broderick+18}.
Nevertheless, even then, a certain fraction of the plasma beam disruption
might still be caused by magnetic fields \cite{Sironi14}, which could
explain the GeV halos of at least some blazars \cite{AlvesBatista+19}.
If magnetic fields really do exist on very large cosmological scales,
they may be primordial in origin.
This may mean that they have been created during or before the
radiation-dominated era of the Universe, for example during one of the
cosmological phase transitions or during inflation.
Inflation was a phase where the conversion from electromagnetic fields
to magnetohydrodynamic fields played an important role, which is what
we are interested in here.

\subsection{Use of Comoving Variables and Conformal Time}

The Universe is expanding with time, as described by the scale factor
$a(t)$.
The equations of magnetohydrodynamics therefore contain additional terms
with factors of $a(t)$ and its time derivatives.
However, by using scaled variables, $\tilde{\AAA}=a\AAA$,
$\tilde{\BB}=a^2\BB$, $\tilde{\EE}=a^2\EE$, $\tilde{\JJ}=a^3\JJ$,
$\tilde{\xx}=\xx/a$, along with conformal time, $\tilde{t}=\int\dd
t/a(t)$, all $a(t)$ factors and other terms involving $a(t)$ disappear
from the magnetohydrodynamic equations \cite{BEO96}.
The velocity is the same in both frames, i.e., $\tilde{\uu}=\uu$.

Given that the equations with tilded variables are equal to the ordinary
ones in a non-expanded Universe, it is convenient to skip all tildes
from now on.
However, when discussing the evolution of the scale factor, for example,
we again need physical time, which will then be denoted by $t_{\rm phys}$,
while $t$ then still denotes conformal time.
Here is where we have a notational dilemma, because in cosmology,
derivatives with respect to physical (or cosmic) time are often denoted
by dots, while those with respect to conformal time are denoted by primes.
We therefore decided here to follow the same convention, so
$a'=\dd a/\dd t$ and $a''=\dd^2 a/\dd t^2$ denote derivatives with
respect to conformal time.

\subsection{Inflationary Magnetogenesis}

Inflationary magnetogenesis models assume the breaking of conformal
invariance through a coupling to a scalar field such as the inflaton.
Another possible coupling is through an axion field, which would result
in helical magnetogenesis, but this will not be considered here.
The dynamics of the scalar field is interesting in its own right;
see Refs.~\cite{Adshead+15, Adshead+16, Adshead+20} for numerical
investigations.
To simplify the model, one commonly replaces this coupling by a prefactor
$f^2$, where $f$ depends on the scale factor of the Universe.
This factor $f^2$ enters in the electromagnetic energy contribution to
the Lagrangian density $f^2 F^{\mu\nu} F_{\mu\nu}$, where $F_{\mu\nu}$
is the Faraday tensor \cite{Ratra92}.
Early approaches to inflationary magnetogenesis exposed specific problems:
the strong coupling and the backreaction problems \cite{Mukhanov09}, as
well as the Schwinger effect constraint, which can lead to a premature
increase in the electric conductivity.
This shorts the electric field and prevents further magnetic field
growth \cite{KA14}.
This is particularly important for models that solve the backreaction
problem by choosing a low energy scale inflation \citep{Ferreira+13},
but could be avoided if charged particles attain sufficiently large
masses by some mechanism in the early Universe \cite{Kobayashi+Sloth19}.
The three problems are avoided by requiring the function $f$ to obey
certain constraints \cite{Sharma+17,Sharma+18}.

Successful models of inflationary magnetogenesis are thus possible,
but this does not mean that the underlying cosmological models are also
physically preferred options.
Nevertheless, for the purpose of discussing the electromagnetic energy
conversion, which is the goal of this paper, those models are a useful
choice.

Three-dimensional simulations of inflationary magnetogenesis have been
performed by assuming an abrupt switch from electromagnetism without
currents and magnetohydrodynamics where the displacement current is
already neglected \cite{BS21, BHS21}.
They solved the evolution equations for the scaled magnetic vector
potential, $\AAAA\equiv f\AAA$, in the Coulomb gauge:
\begin{equation}
\left[\frac{1}{c^2}\frac{\partial^2}{\partial t^2}
-\nabla^2-k_*^2(t)\right]\AAAA=0.
\label{dA2dt2}
\end{equation}
where $k_*^2(t)=f''/f$ is a generation term, because it destabilizes
the field at large length scales for wavenumbers $k<k_*(t)$.
Analogous to the primes on $a(t)$, primes on $f(t)$ also denote conformal
time derivatives.
Toward the end of the reheating phase, where $f\to1$, we expect
$k_*(t)\to0$.

Our aim here is to present calculations where the transit from vacuum
to high conductivity is continuous.
In particular, to calculate the generation term $k_*^2(t)$, one
commonly uses a power law representation in terms of $a(t)$ of
the form $f\propto a^{\alpha}$ with $\alpha>0$ during inflation
and $f\propto a^{-\beta}$ with $\beta>0$ during reheating \cite{Sub10}.
We are here only interested in the reheating phase where
$a(t)\propto t^2$ \cite{Sharma+17,Sharma+18} such that it is
unity when the radiation-dominated era begins, and therefore
$f=1$ and $k_*^2(t)=0$ for $a>1$.
For $a<1$, by contrast, we have
\begin{equation}
k_*^2(t)=\beta\left[(\beta+1)(a'/a)^2-a''/a\right].
\end{equation}
Note that for $a=t^2$, we have $a'=2t$ and $a''=2$,
so $(a'/a)^2=4/t^2$ and $a''/a=2/t^2$, and therefore
$f''/f=2\beta(2\beta+1)/t^2$.

Contrary to the earlier numerical work \cite{BS21, BHS21}, the
displacement current is now included at all times.
However, there is still a problem in that $k_*^2(t)$
has a discontinuity from $k_*^2(1)=\beta(\beta+1)\neq0$ to zero
at the moment when the conductivity is turned on.
In the simulations, this did not seem to have any serious effect on the
results, because the magnetic field at the end of the electromagnetic
phase only acted as an initial condition for the magnetohydrodynamic
calculation after the switch.
In a continuous calculation without switch, however, this problem
must be avoided.
This will be addressed next.

\subsection{Continuous Version of the Generation Term}

An instructive way of obtaining a smooth transition from a quadratic to
a linear growth profile of $a(t)$ is obtained by solving the Friedmann
equations for a piecewise constant equation of state, $w(a)$, which
relates the pressure with the density through $p=w\rho$.
Under the assumption of zero curvature, i.e., the Universe is conformally
flat, but expanding, the Friedmann equations can be written as a single
equation which, in physical time, takes the form
\begin{equation}
a^{-1}\,\dd^2a/\dd t_{\rm phys}^2
=-\half H^2\left[1+3w(a)\right],
\label{Friedmann_phys}
\end{equation}
where $H=a^{-1}\dd a/\dd t_{\rm phys}$ is the standard Hubble parameter.
Here, $w(a)=1/3$ during the radiation-dominated era and $w(a)=0$ during
reheating when there were no photons, which is therefore equivalent to the
matter-dominated era that also occurs later after recombination and
before the Universe began to accelerate again.
The accelerated exponential expansion of the Universe during inflation,
and also the late acceleration of the present Universe, correspond to
$w=-1$, but this will not be considered in the present paper.

It is convenient to solve the Friedmann equation with zero curvature
in conformal time.
It then takes the form $a''/a=\half{\cal H}^2\,(1-3w)$,
where ${\cal H}=a'/a$ is the conformal Hubble parameter.
It is related to the usual one, $H$, through
${\cal H}=\dd a/\partial t_{\rm phys}=aH$.
Note the opposite sign of the terms on the right-hand side and the
opposite sign in front of $3w(a)$ compared to the formulation in terms
of physical time.
The equation for $a''$ is easily solved by splitting it into two
first-order equations and introducing a new variable $b(t)$ and solving
for
\begin{equation}
a'=b, \quad b'=(b^2/2a)(1-3w);
\end{equation}
see also Ref.~\cite{HRPB23} for similar work in another context.

\begin{figure}[t!]
\hspace{-16pt}\includegraphics[width=\columnwidth]{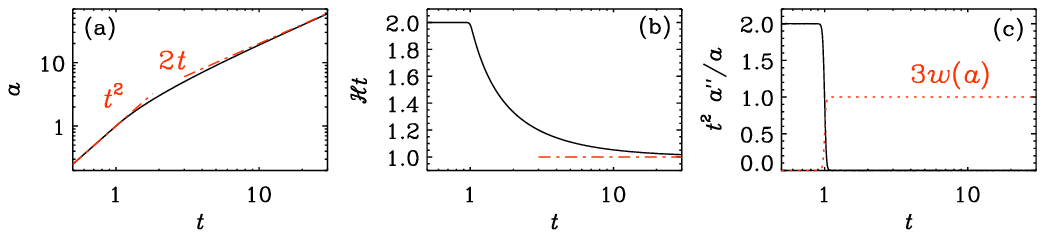}
\caption[]{
$t$ dependence of
(\textbf{a})~the scale factor $a(t)$,
(\textbf{b})~the compensated Hubble coefficient ${\cal H}t=t\,a'/a$, and
(\textbf{c})~the compensated left-hand side of the Friedmann equation, $t^2\,a''/a$.
In (\textbf{a}), the asymptotic dependences $a=t^2$ and $2t$ for $t\ll1$
and $\gg1$ are overplotted as dashed-dotted orange lines.
In (\textbf{c}), the function $3w(a(t))$
is overplotted as a dotted red line.
\label{pprof}}\end{figure}

\Fig{pprof} shows the solution for $a(t)$ and the ratios
$a'/a$ and $a''/a$ compensated by $t$ and $t^2$, respectively,
which allows us to see more clearly how $a'/a$ changes from $2/t$
to $1/t$ and $a''/a$ changes from $2/t^2$ to zero as we go from the
reheating era to the radiation-dominated Universe after reheating.

\begin{figure}[t!]
\hspace{-16pt}\includegraphics[width=.9\columnwidth]{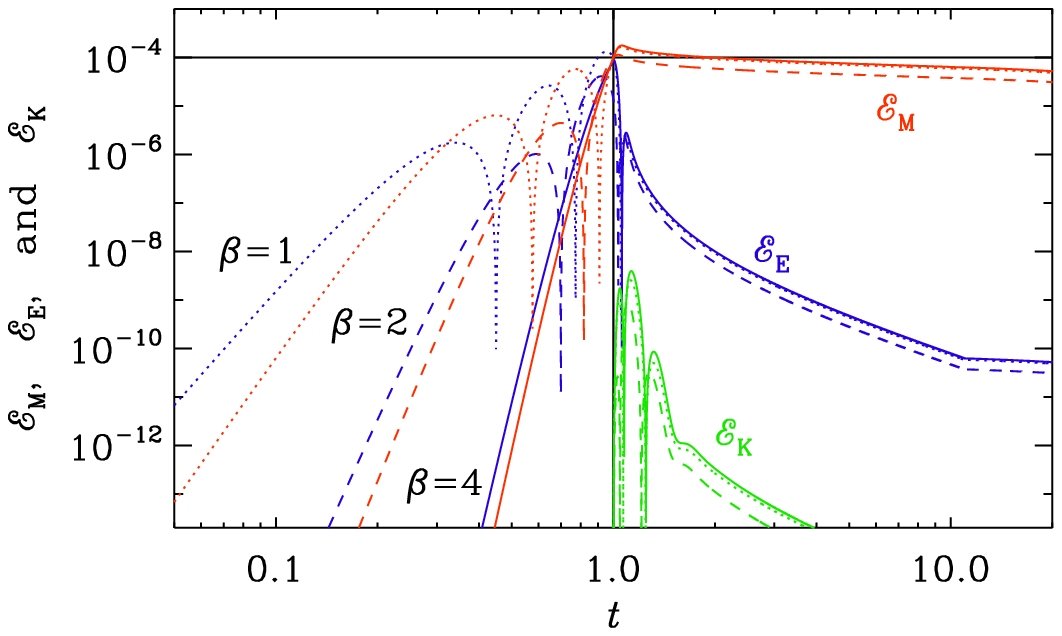}
\caption[]{
$t$ dependence of $\EEM$ (red), $\EEEl$ (blue), and $\EEK$ (green)
for runs with $\beta=1$ (dotted lines), $2$ (dashed lines),
and $4$ (solid lines) for Set~(i) with $k=10$, $t_0=1$, and $t_{\rm trans}=10$.
The initial amplitudes have been arranged such that $\Brms=0.01$ at $t=1$.
From the double-logarithmic representation, we see that the
growth of $\EEM$ and $\EEEl$ is algebraic, and much faster
for the models with a larger value of $\beta$.
Before $t=1$, $\EEEl$ dominates over $\EEM$, but drops immediately
after $t=1$, when resistivity emerges and kinetic energy is being
generated.
Both $\EEM$ and $\EEK$ are larger for larger values of $\beta$.
\label{pcomp_eb_toff1_k10_nocurlyA}}\end{figure}

The generation term $k_*^2(t)\equiv f''/f$
determines the wavenumber below which the solution is still unstable.
However, since $k_*^2(t)=2\beta(2\beta+1)/t^2$, we have
$ctk=\const\approx2\beta+1/2$; see \Tab{Tbet}.
In \Fig{pcomp_eb_toff1_k10_nocurlyA}, we plot the evolution of $\EEEl$,
$\EEM$, and $\EEK$ for $k=10$ for all three values of $\beta$: 1, 2,
and 4.
Here and below, the initial amplitudes have been arranged such that
$\EEM=10^{-4}$ at $t=t_0$.
In all cases, the solution has become stable by the
time $t=t_0=1$, and we see electromagnetic oscillations toward the
end of the reheating phase before conductivity turns on at $t_0=1$.
This is here referred to as Set~(i).


\begin{specialtable}[H]\caption{
Parameters relevant for the models with different values of $\beta$.
\label{Tbet}}
\begin{tabular}{lccccccc}
\toprule
$\beta$ & $2\beta+1$ & $2\beta+1/2$ & $k_*(1)$ \\
\midrule
1 & 3 & 2.5 & 2.45 \\
2 & 5 & 4.5 & 4.47 \\
4 & 9 & 8.5 & 8.49 \\
\bottomrule
\end{tabular}
\end{specialtable}

It is easy to see that on large length scales, when the $\nabla^2$
operator in \Eq{dA2dt2} is negligible compared with $k_*^2(t)$,
we have
\begin{eqnarray}
&{\cal A}_z(x,t)=A_0 t^{2\beta+1}k^{-1}\cos kx,\quad
A_z(x,t)={\cal A}_z/f=A_0 t^{\beta+1}k^{-1}\cos kx,\\
&B_y(x,t)=A_0 t^{\beta+1}\sin kx,\quad
E_z(x,t)=-\partial A_z/\partial t=-(\beta+1) A_0 t^{\beta}k^{-1}\cos kx.
\end{eqnarray}
Thus, for $ckt\ll1$, corresponding to super-horizon scales, where and
when the modes are still unstable, we have $t\Erms/\Brms\approx\beta+1$.
On smaller length scales, i.e., for larger $k$ values, the modes become
stable and we have the usual electromagnetic waves.

\begin{figure}[t!]
\hspace{-16pt}\includegraphics[width=.9\columnwidth]{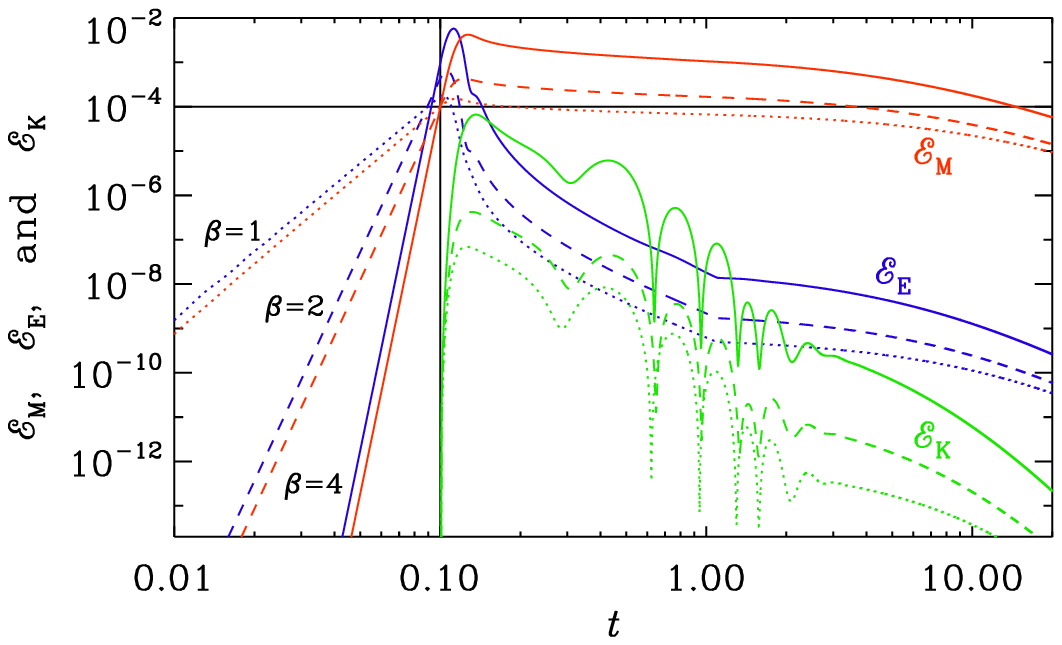}
\caption[]{
Similarly to \Fig{pcomp_eb_toff1_k10_nocurlyA}, but for
Set~(ii) with $k=10$, $t_0=0.1$, and $t_{\rm trans}=1$.
\label{pcomp_eb_nocurlyA}}\end{figure}

\begin{figure}[b!]
\hspace{-16pt}\includegraphics[width=.9\columnwidth]{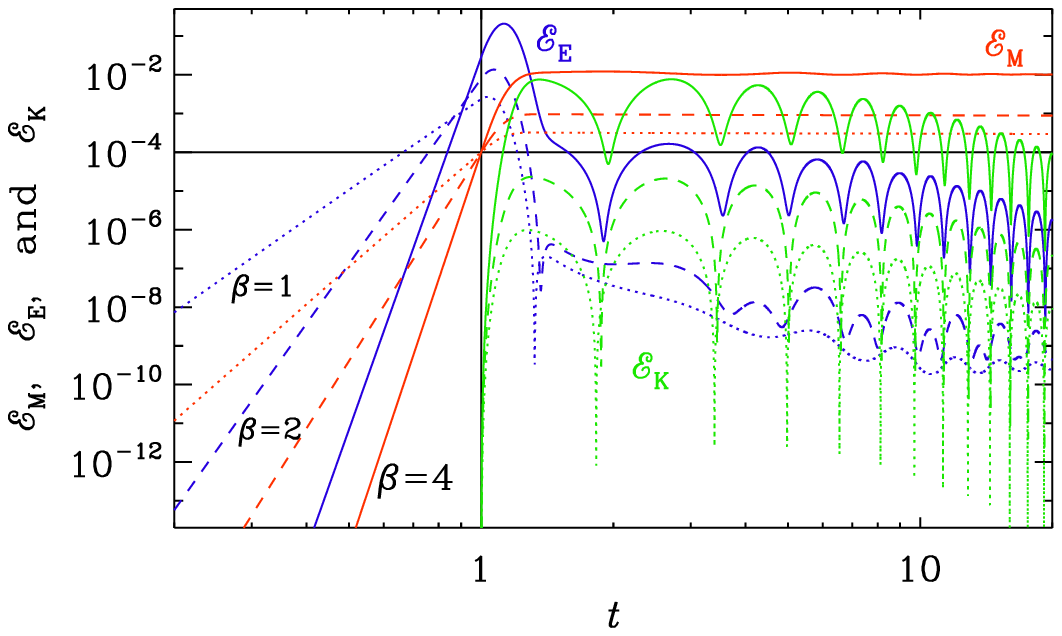}
\caption[]{
Similarly to \Fig{pcomp_eb_toff1_k10_nocurlyA}, but for
Set~(iii) with $k=1$, $t_0=1$, and $t_{\rm trans}=10$.
\label{pcomp_eb_toff1_nocurlyA}}\end{figure}

When modeling the transition from a vacuum to that of high conductivity
and the corresponding Joule heating, we still need to make a choice as
to when $\sigma$ would begin to increase, i.e., we need to choose values
of $t_0$ and $t_{\rm trans}$.
If we choose the value of $t_0$ to be too large, we obtain solutions
where electromagnetic waves have already been established; see
\Fig{pcomp_eb_toff1_k10_nocurlyA}.
The smallest wavenumber in our one-dimensional domain is $k=10$, so
by the time $t=1$, even the largest modes in the domain are stable.
We also see that at early times, $\EEEl$ and $\EEM$ grow in an algebraic
fashion and then become oscillatory when $k_*(t)$ has dropped below $k$.
At $t=t_0=1$, when conductivity turns on, the electric energy decreases
rapidly, while the magnetic energy diminishes only very slowly.
The generated hydrodynamic energy is however small.
This is similar to what we studied in \Sec{Transition}.

Our objective here is to study cases that are different
from what was studied in \Sec{Transition}.
Therefore we now choose Set~(ii) with $t_0=0.1$ and $t_{\rm trans}=1$
(\Fig{pcomp_eb_nocurlyA}) and another Set~(iii) with $k_1=1$ and $t_0=1$
(\Fig{pcomp_eb_toff1_nocurlyA}).
Again, the electric energy drops significantly when conductivity turns on,
but now there is a much larger spread in the resulting maximum magnetic
energies for the three cases with $\beta=1$, 2, and 4.
For $\beta=4$, $\EEK$ reaches about one percent of $\EEM$ at $t=0.2$,
for example.





When increasing the wavenumber to $k=1$, the largest modes are
still unstable for the three cases with $\beta=1$, 2, and 4; see
\Fig{pcomp_eb_toff1_nocurlyA}.
Here, $k=1$ and $t_0=1$ and $t_{\rm trans}=10$.
The spread in the magnetic energy is similar, but the maximum kinetic
energy is now much larger; see \Tab{Tbet2}.

\begin{specialtable}[b!]
\caption{Summary of various extrema for each of the three sets of models
and values of $\beta$.
\label{Tbet2}}
\begin{tabular}{cccllll}
\toprule
Set & $k$ & $t_0$ & variable & $\beta=1$ & $\beta=2$ & $\beta=4$ \\
\midrule
  (i) & 10 & 1 &$\max{\cal E}_{\rm E}$&$1.3\times10^{-4}$&$4.1\times10^{-5}$&$8.3\times10^{-5}$\\
 (ii) & 10 &0.1& &$1.8\times10^{-4}$&$6.4\times10^{-4}$&$5.8\times10^{-3}$\\
(iii) &  1 & 1 & &$2.7\times10^{-3}$&$1.4\times10^{-2}$&$2.1\times10^{-1}$\\
\midrule
  (i) & 10 & 1 &$\max{\cal E}_{\rm M}$&$1.6\times10^{-4}$&$1.1\times10^{-4}$&$1.8\times10^{-4}$\\
 (ii) & 10 &0.1& &$1.5\times10^{-4}$&$4.3\times10^{-4}$&$4.2\times10^{-3}$\\
(iii) &  1 & 1 & &$3.2\times10^{-4}$&$9.6\times10^{-4}$&$1.2\times10^{-2}$\\
\midrule
  (i) & 10 & 1 &$\max{\cal E}_{\rm K}$&$2.5\times10^{-9}$&$9.0\times10^{-10}$&$3.9\times10^{-9}$\\
 (ii) & 10 &0.1& &$6.9\times10^{-8}$&$4.2\times10^{-7}$&$6.6\times10^{-5}$\\
(iii) &  1 & 1 & &$9.8\times10^{-7}$&$2.3\times10^{-5}$&$7.7\times10^{-3}$\\
\midrule
  (i) & 10 & 1 &$\max Q_{\rm G}$&$1.4\times10^{-4}$&$2.0\times10^{-4}$&$1.4\times10^{-3}$\\
 (ii) & 10 &0.1& &$4.4\times10^{-3}$&$3.8\times10^{-2}$&$8.6\times10^{-1}$\\
(iii) &  1 & 1 & &$8.4\times10^{-3}$&$7.8\times10^{-2}$&$2.7\times10^{0}$\\
\midrule
  (i) & 10 & 1 &$\max(-Q_{\rm E})$&$2.0\times10^{-3}$&$9.3\times10^{-4}$&$2.0\times10^{-3}$\\
 (ii) & 10 &0.1& &$5.3\times10^{-3}$&$2.2\times10^{-2}$&$2.7\times10^{-1}$\\
(iii) &  1 & 1 & &$1.3\times10^{-3}$&$4.7\times10^{-3}$&$5.9\times10^{-2}$\\
\midrule
  (i) & 10 & 1 &$\max\epsM$&$2.0\times10^{-3}$&$4.1\times10^{-4}$&$2.5\times10^{-3}$\\
 (ii) & 10 &0.1& &$2.1\times10^{-2}$&$1.1\times10^{-1}$&$1.5\times10^{0}$\\
(iii) &  1 & 1 & &$3.5\times10^{-2}$&$2.5\times10^{-1}$&$5.8\times10^{0}$\\
\midrule
  (i) & 10 & 1 &$\max(-\dot{\cal E}_{\rm E})$&$3.3\times10^{-3}$&$6.3\times10^{-4}$&$2.7\times10^{-3}$\\
 (ii) & 10 &0.1& &$1.6\times10^{-2}$&$5.9\times10^{-2}$&$5.6\times10^{-1}$\\
(iii) &  1 & 1 & &$2.4\times10^{-2}$&$1.3\times10^{-1}$&$1.9\times10^{0}$\\
\midrule
  (i) & 10 & 1 &$\max\dot{\cal E}_{\rm E}$&$1.2\times10^{-3}$&$3.4\times10^{-4}$&$8.0\times10^{-4}$\\
 (ii) & 10 &0.1& &$8.6\times10^{-3}$&$5.4\times10^{-2}$&$5.4\times10^{-1}$\\
(iii) &  1 & 1 & &$1.8\times10^{-2}$&$1.3\times10^{-1}$&$2.0\times10^{0}$\\
\bottomrule
\end{tabular}
\end{specialtable}

\subsection{Energy Conversions during Reheating}
\label{EnergyConversionsReheating}

\begin{figure}[t!]\begin{center}
\includegraphics[width=\columnwidth]{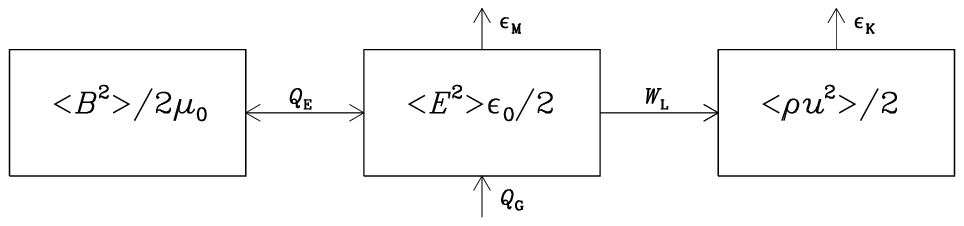}
\end{center}\caption[]{
Similar to \Fig{pboxes_BEU}, but now with inflationary
magnetogenesis energy generation and energy exchange between
electric and magnetic energies in both directions.
}\label{pboxes_G}\end{figure}

During reheating, there is an additional source of energy resulting from
the generation term $k_*^2(t)$.
The term $k_*^2(t)$ appeared in \Eq{dA2dt2} for $\AAAA=f\AAA$.
However, to write down the relevant equation for
$\EE=-\partial\AAA/\partial t$, we have to revert to the original
equation for $\AAA$, which reads \cite{Sub10}
\begin{equation}
\left[\frac{1}{c^2}\left(\frac{\partial^2}{\partial t^2}
+2\frac{f'}{f}\frac{\partial}{\partial t}\right)
-\nabla^2\right]\AAA=0.
\label{nocurlyA}
\end{equation}
Thus, \Eq{dEdt} with the current density term restored, now becomes
\begin{equation}
\frac{1}{c^2} \left(\frac{\partial\EE}{\partial t}
+2\frac{f'}{f}\EE\right)=\nab\times\BB-\mu_0\JJ,
\label{dEdt_infl}
\end{equation}
and therefore, \Eq{dE2dt2} for the electric energy
now has an extra term and reads
\begin{equation}
\frac{\dd}{\dd t}\bbra{\epsilon_0\EE^2/2}=-2(f'/f)\bbra{\epsilon_0\EE^2}
+\bbra{\EE\cdot\nab\times\BB/\mu_0}-\bbra{\JJ\cdot\EE}.
\label{dE2dt3}
\end{equation}
During reheating with $f\propto a^{-\beta}\propto t^{-2\beta}$, we have
$f'/f=-2\beta/t$, so the first term on the right-hand side of \Eq{dE2dt3}
is positive for $\beta>0$, so there is growth of the electric energy.
Similarly to what was performed in \Sec{Transition},
we can write the electric energy equation more compactly as
$\dot{\cal E}_{\rm E}=Q_{\rm G}+Q_{\rm E}-\epsM-W_{\rm L}$, where
$Q_{\rm G}=-4(f'/f)\EEEl$ is now the dominant source, but here $Q_{\rm E}$
plays the role of a sink during the first part of the evolution.
This equation generalizes \Eq{epsM_eqn} to the case with electromagnetic
field generation during reheating; see also \Fig{pboxes_G}.
Unlike the earlier case of \Fig{pboxes_BEU}, where there was no energy
input, we here have a system that it driven by energy input through the
$Q_{\rm G}$ term.

\begin{figure}[t!]
\hspace{-16pt}\includegraphics[width=.9\columnwidth]{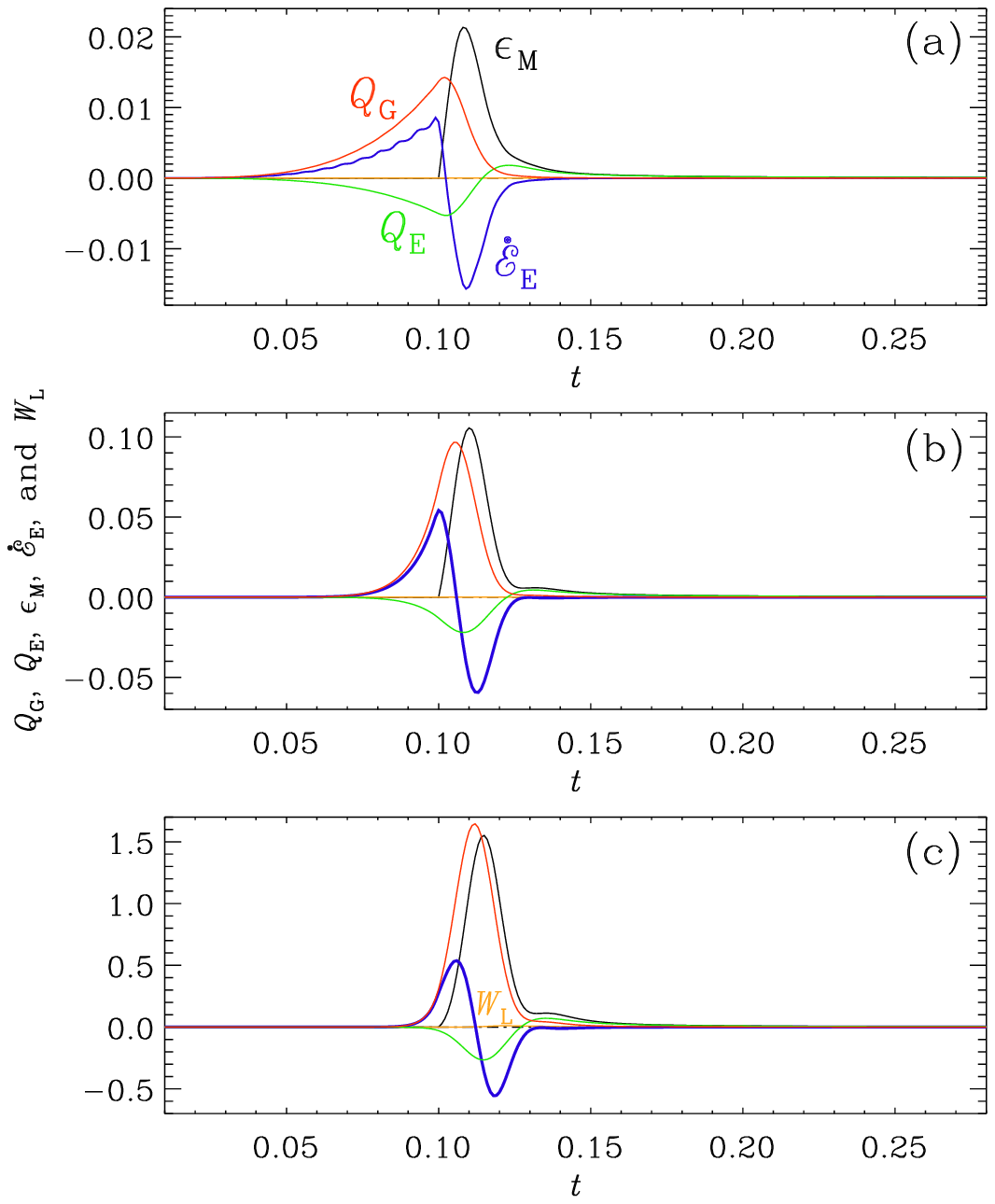}
\caption[]{
$t$ dependence of $Q_{\rm G}$ (red), $Q_{\rm E}$ (green), $\epsM$ (black),
$\dot{\cal E}_{\rm E}$ (blue), and $W_{\rm L}$ (orange) for the runs of
Set~(ii) in \Fig{pcomp_eb_nocurlyA} with
(\textbf{a}) $\beta=1$, (\textbf{b}) $\beta=2$, and
(\textbf{c}) $\beta=4$ for $t_0=0.1$ and $t_{\rm trans}=1$.
\label{pcomp_infl_nocurlyA}}\end{figure}

\begin{figure}[t!]
\hspace{-16pt}\includegraphics[width=.9\columnwidth]{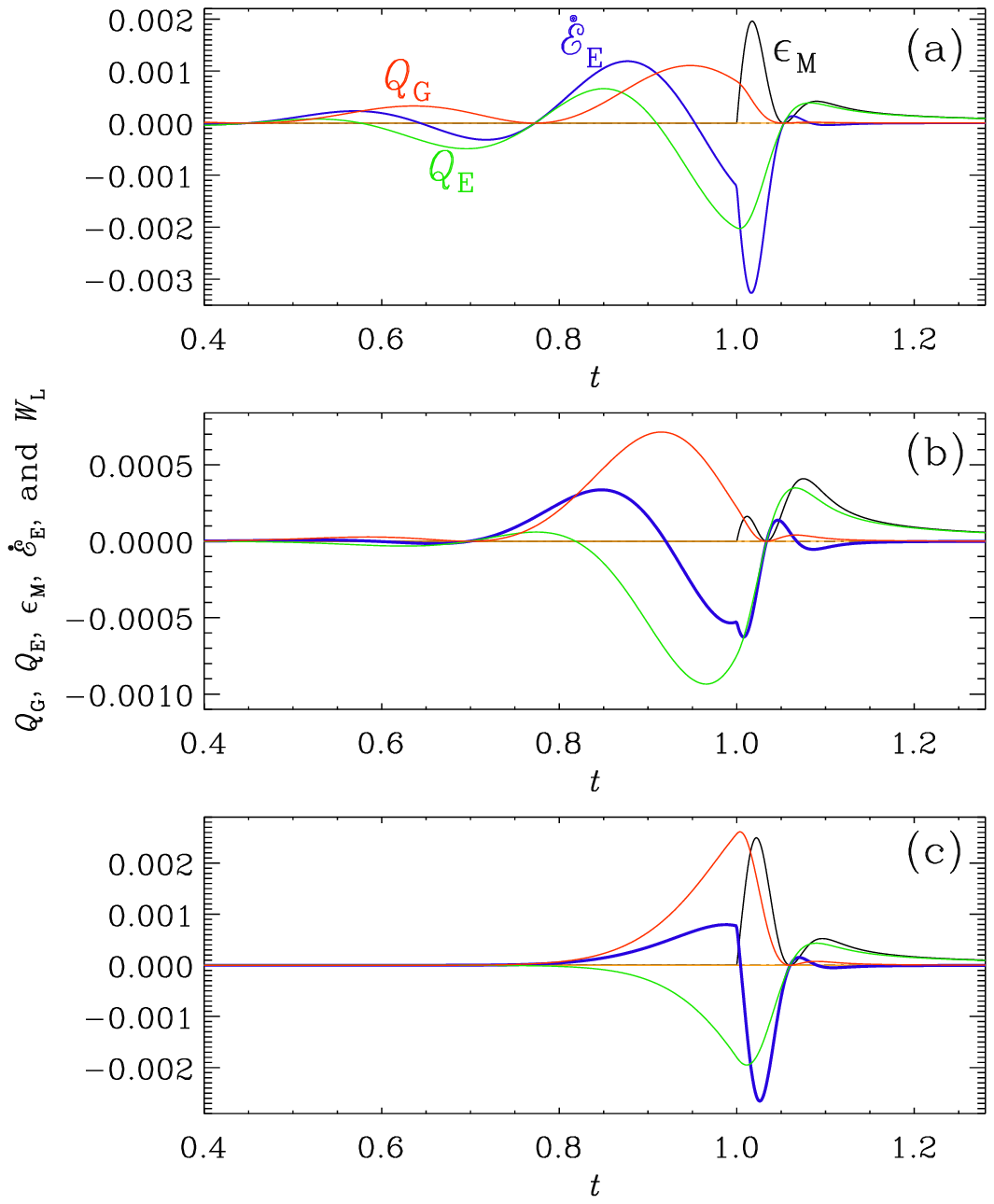}
\caption[]{
Similarly to \Fig{pcomp_infl_nocurlyA}, but for the runs of Set~(i)
in \Fig{pcomp_eb_toff1_k10_nocurlyA} with $t_0=1$ and $t_{\rm trans}=10$.
Note that $Q_{\rm G}$ and $Q_{\rm E}$ vary in anti-phase.
\label{pcomp_infl_toff1_k10_nocurlyA}}\end{figure}

The evolution of $Q_{\rm G}$, $Q_{\rm E}$, $\dot{\cal E}_{\rm E}$,
and $\epsM$ is shown in \Fig{pcomp_infl_nocurlyA} during magnetic field
generation in case (ii) for all three values of $\beta$.
It is instructive to write the electric energy equation as
\begin{equation}
Q_{\rm G}=\epsM+\dot{\cal E}_{\rm E}+W_{\rm L}-Q_{\rm E}.
\end{equation}
Comparing the three panels of \Fig{pcomp_infl_nocurlyA}, we thus see that for
$\beta=1$, there is a slow generation phase starting much before $t_0$.
It should be noted, however, that the ranges on the vertical axes are
different for the different panels.

At $t=t_0=0.1$, there is a rise of conductivity, and therefore a sharp
rise in Ohmic heating, $\epsM$.
This is also the time when $\dot{\cal E}_{\rm E}$ reaches a maximum and
becomes negative shortly thereafter.
For large values of $\beta$, this moment happens a bit later, at $t=0.11$
compared to $t=0.10$ for $\beta=1$.
Note that, while for $\beta=4$ the maxima of $Q_{\rm G}$ and $\epsM$
are similar, for smaller values of $\beta$, the maxima of $\epsM$ are
much larger than those of $Q_{\rm G}$.
Instead, for $\beta=1$, for example, we have
$\epsM\approx-\dot{\cal E}_{\rm E}$, i.e., almost the entire heating
is here caused by dissipation of electric energy.

In \Fig{pcomp_infl_toff1_k10_nocurlyA}, we show a plot similar to
\Fig{pcomp_infl_nocurlyA}, but for the case (i), where all modes were
already oscillatory at $t=t_0=1$, when conductivity turned on.
The Ohmic heating now plays a minor role in the sense that its maximum
value is much less than the extrema of $Q_{\rm G}$, $Q_{\rm E}$, and
$\dot{\cal E}_{\rm E}$.
For $\beta=1$ and 2, we see that $\dot{\cal E}_{\rm E}$ and $Q_{\rm E}$
are nearly in phase shortly before conductivity turns on.
This means that the electric and magnetic energies are strongly coupled
and a flow of energy from magnetic to electric energy ($Q_{\rm E}>0$)
leads to an increase in electric energy ($\dot{\cal E}_{\rm E}>0$).
This is expected, because there is only an oscillatory exchange between
electric and magnetic energies.
For $\beta=4$, on the other hand, the oscillatory phase just started to
develop shortly before $t=1$, but the curves are similar to those for
$\beta=1$ and 2, although shifted toward earlier times.
The time of the first maximum of $Q_{\rm G}$ is at $t=0.9$ for $\beta=4$,
while for $\beta=2$, it is at $t=0.46$ and for $\beta=1$ it is at
$t=0.25$, and we see that the profiles of all curves are indeed very
similar around those times.



\section{Conclusions}
\label{Conclusions}

In this paper, we have studied the conversion of electromagnetic energy
into kinetic and thermal energies as the electric conductivity transits
from zero (vacuum) to large values.
This problem has relevance to the reheating phase at the end of
cosmological inflation and before the emergence of a relatively long
radiation-dominated era before the time of recombination, which is
much later (on a logarithmic time scale).
While not much is known about the physical processes leading to reheating
and the emergence of conductivity, a lot can now be said about the
general process of such an energy conversion.

Already in the absence of cosmological expansion, we have seen that the
transition to conductivity involves an oscillatory exchange between
electric and magnetic energies.
It is mainly the electric energy reservoir that delivers energy to the
kinetic and thermal energy reservoirs, and not the magnetic energy directly,
as in magnetohydrodynamics.
We knew already from earlier work that the duration of the transit plays
a significant role in causing a drop in magnetic energy.
We now also see that this drop depends on the magnetic field strength
and thus the typical Alfv\'en speed.
The drop can become small if the Alfv\'en speed becomes comparable to
the speed of light.
Furthermore, for short transits, we have seen that energy transfer
between electric and magnetic energies is small and that the initial
electric energy goes directly into thermal energy.
For longer transits, however, the mutual exchange with magnetic energy
becomes approximately equal to the thermal energy loss, so thermalization
now also involves the magnetic energy reservoir.

When applying electromagnetic energy conservation to the problem of
reheating, we have a new quality in the model in that there is now also
energy transfer through conformal invariance breaking, which may occur
during inflation and reheating.
This is obviously speculative \cite{Ratra92, Kronberg94, Widrow02,
Giovannini04, Kandus+11, DN13, Sub16, Vachaspati21}, but a very promising
scenario for the generation of large-scale magnetic fields in the early
Universe and for explaining the observed lower limits of the intergalactic
magnetic field on megaparsec length scales \cite{NV10, Taylor+11}.

The present study has shown that significant work can be done by the Lorentz
force when the electromagnetic energy conversion happens early and on
scales large enough so that the modes are still growing in time.
This is because there is then significant excess of electric energy
over magnetic.
This is an effect that was ignored in previous simulations of
inflationary magnetogenesis and, in particular, in studies of the
additional contributions to the resulting relic gravitational wave
production \cite{BS21, BHS21}.

It will be useful to extend our studies to turbulent flows and magnetic
fields. This requires that one solves for the evolution of $\rho_{\rm e}$
and that $\nab\cdot\EE=\rho_{\rm e}/\epsilon_0$ is obeyed at all times.
This constraint was automatically obeyed in our one-dimensional models.
In future, it would also be interesting to study
dynamo action in situations of moderate magnetic conductivity where coupling
with the electric energy reservoir could reveal new aspects.

\vspace{6pt}
\authorcontributions{Conceptualization, A.B.\ and N.N.P.; methodology, A.B.; software, A.B.\ and N.N.P.
All authors have read and agreed to the published version of the manuscript.}

\funding{This research was funded by Vetenskapsr{\aa}det grant number 2019-04234 and NASA ATP award number 80NSSC22K0825.}

\dataavailability{
The source code used for the simulations of this study,
the {\sc Pencil Code} \citep{JOSS}, is freely available on
\url{https://github.com/pencil-code/} (accessed on 1 August 2023).
The DOI of the code is \url{https://doi.org/10.5281/zenodo.2315093} (accessed on 1 August 2023).
The simulation setups and the corresponding secondary data are available on\\
\url{http://norlx65.nordita.org/~brandenb/projects/EMconversion} (accessed on 1 August 2023)
and on \url{https://doi.org/10.5281/zenodo.8203242} (accessed on 1 August 2023).
} 

\acknowledgments{The authors thank Johan Anderson for suggesting this contribution
to the special issue on Energy Transfer and Dissipation in Plasma Turbulence.
We are also indebted to the comments and suggestions of the two referees,
which have helped to improve the presentation.
We acknowledge the allocation of computing resources provided by the
Swedish National Allocations Committee at the Center for Parallel
Computers at the Royal Institute of Technology in Stockholm and
Link\"oping.}

\conflictsofinterest{The authors declare no conflicts of interest.}

\end{paracol}
\reftitle{References}
\externalbibliography{yes}
\bibliography{ref}

\end{document}